\documentclass[useAMS]{mn2e}
\usepackage{graphicx,amssymb,amsmath}
\newcommand{\scs}{\scriptsize}

\title[NGC 2527, NGC 2682, NGC 2482, NGC 2539, NGC 2335, NGC 2251, NGC 2266]
  {Comprehensive abundance analysis of red giants in the open clusters NGC 2527, 2682,
    2482, 2539, 2335, 2251 and 2266 }
\author[A. B. S. Reddy, S. Giridhar and D. L. Lambert ]
  { Arumalla B.S. Reddy$^1$\thanks{E-mail: sudha@iiap.res.in (ABSR);
giridhar@iiap.res.in (SG); dll@astro.as.utexas.edu (DLL)},
    Sunetra Giridhar$^1$ and David L. Lambert$^2$  \\
  $^1$Indian Institute of Astrophysics, Bangalore 560034, India \\
  $^2$W.J. McDonald Observatory, The University of Texas at Austin, Austin, TX 78712
- 0259, USA }

\begin{document}

\pagerange{\pageref{firstpage}--\pageref{lastpage}} \pubyear{2013}

\maketitle

\label{firstpage}

\begin{abstract}

We have analyzed high-resolution echelle spectra of red giant members for seven open 
clusters in the Galactic anticentre direction to explore their chemical compositions. 
Cluster membership has been confirmed by radial velocity. The 
spread in temperatures and gravities being very small among the red giants, nearly the 
same stellar lines were employed for all stars 
thereby reducing the abundance errors: the errors of the average 
abundance for a cluster were generally in the 0.02 to 0.05 dex range. Our present
sample 
covers Galactocentric distances of 8.3 to 11.3 kpc and an age range of 0.2 to 4.3 Gyr.
A careful comparison of our results for the cluster NGC 2682 (M 67) to other high-resolution 
abundance studies in the literature shows general good agreement for almost all elements
in common.

\end{abstract}


\section{Introduction} 

This is our second paper reporting abundance measurements for red giants in
open clusters (OCs) lacking detailed information on their chemical
composition. 
In our first paper (Reddy et al. 2012, hereafter paper {\scs I}) we have 
presented  abundance measurements for four
open clusters (OCs) whose Galactocentric distances (R$_{\rm gc}$)
lie between 8.3 and 10.5 kpc, where the Sun's Galactocentric distance
is taken as 8.0$\pm0.6$ kpc 
 (Ghez et al. 2008). Here, we add seven OCs six of which have not previously
analysed in detail also in the Galactic anticentre
direction in the disc.
Our overall goal is to improve understanding of 
Galactic chemical evolution. 

The abundance variation of chemical elements with Galactocentric 
distance (R$_{gc}$) (i.e., the radial metallicity gradient) along (and across)
the disc and the gradient's 
temporal variation over the disc's lifetime  put  constraints
on Galactic chemical evolution models which are controlled, in large part,
by  variations of such quantities as 
the initial mass function (IMF), the star formation rate (SFR), gas flows
in and out of the disc as well as through the disc.
Particularly informative diagnostics are the spatial and temporal variations
of relative abundances which are traditionally referenced to the iron abundance,
i.e., ratios such as O/Fe which primarily samples the relative
rates of Type II and Type Ia supernovae and Ba/Fe where Ba measures heavy
element synthesis by AGB stars.

Abundance variations across the Galaxy, particularly measurements of the
radial gradient, have been measured using a wide variety of objects including
H\,{\sc ii} regions, hot young stars, Cepheid variables, planetary nebulae, red giants,
and OCs. Various disagreements and inconsistencies remain - see, for example,
Magrini et al. (2010). OCs either through spectroscopy of their red giants
or cooler main sequence stars provide not only abundance estimates for many
elements -- essentially, elements sampling all the major processes of
stellar nucleosynthesis -- but a collection of stars with a well determined age,
distance and metallicity. Moreover, it is possible to estimate from an OC's
space motion and a model of the Galactic gravitational potential the birthplace
of the OC.

OCs dissolve over time thanks to tidal forces and feed stars
into the field.  Intracluster
abundance variations, especially in the $\alpha$- and the heavy elements provide,
in principle, a way to 
chemically tag  groups of field stars to their parent cluster 
(Freeman \& Bland-Hawthorn, 2002; De Silva et al. 2006, 2007). 
Such studies benefit from the largest possible sample of OCs
with all properties analysed homogeneously  with high-resolution spectroscopy.  

Addressing abundance variations across the Galactic disk and applying
chemical tagging requires homogeneous and accurate abundance analyses
of as large a sample of OCs as possible. In this paper, we continue our modest
efforts in this regard by extending
our homogeneous abundance analysis for a sample of seven OCs  
for many elements including the s- and r-process elements for which the 
measurements are often lacking in the literature.

The layout of the present paper is as follows: 
In Section 2 we describe the data selection, observations and data reduction, and
Section 3 is
devoted to the abundance analysis. In Section 4 we present our results and compare
them with the abundances derived from samples of field giants.
Finally, in Section 5 we give the conclusions.
 
\begin{table*}
 \centering
  \caption{Target clusters and their properties from the literature.}
 \label{tab1}
\begin{tabular}{lcccccccl}
  \hline
\multicolumn{1}{l}{Cluster}&  
\multicolumn{1}{c}{$\ell$}&  
\multicolumn{1}{c}{b}&  
\multicolumn{1}{c}{Age}&  
\multicolumn{1}{c}{[Fe/H]$_{\rm phot.}$}&  
\multicolumn{1}{c}{R$_{\rm gc}$}&
\multicolumn{1}{c}{(m-M)$_{V}$}&
\multicolumn{1}{c}{E(B-V)}&
\multicolumn{1}{c}{[Fe/H]$_{\rm ref}$} \\
\multicolumn{1}{c}{}&
\multicolumn{1}{c}{(deg.)}&
\multicolumn{1}{c}{(deg.)}&
\multicolumn{1}{c}{(Gyr)}&
\multicolumn{1}{c}{(dex.)}&
\multicolumn{1}{c}{(kpc)}&
\multicolumn{1}{c}{(mag.)}&
\multicolumn{1}{c}{(mag.)}&
\multicolumn{1}{l}{ } \\
\hline
NGC 2251 & 203.58 & $+$00.10 & 0.27 & $-$0.20 &  9.2 & 11.19 & 0.19 & Parisi et al.
(2005)  \\
NGC 2266 & 187.79 & $+$10.29 & 1.20 & $-$0.39 & 11.3 & 12.97 & 0.10 & Kaluzny et al.
(1991) \\
NGC 2335 & 223.60 & $-$01.18 & 0.16 & $-$0.03 &  9.1 & 11.98 & 0.39 & Twarog et al.
(1997)  \\
NGC 2482 & 241.62 & $+$02.03 & 0.40 & $+$0.12 &  8.7 & 10.93 & 0.09 & Twarog et al.
(1997)  \\
NGC 2527 & 246.08 & $+$01.85 & 0.45 & $-$0.09 &  8.3 & 09.01 & 0.04 & Piatti et al.
(1995)  \\
NGC 2539 & 233.70 & $+$11.11 & 0.37 & $-$0.20 &  8.9 & 10.93 & 0.08 & Claria \&
Lapasset (1986) \\
NGC 2682 & 215.69 & $+$31.89 & 4.30 &    0.00 &  8.6 &  9.97 & 0.06 & Dinescu et al.
(1995)  \\

\hline
\end{tabular}
\end{table*}

\begin{table*}
 \centering
 \caption{ The observed stars } 
 \label{tab2}
\begin{tabular}{lccccccccl}
  \hline
\multicolumn{1}{l}{Cluster}& \multicolumn{1}{c}{Star ID}&  
\multicolumn{1}{c}{$\alpha(2000.0)$}& \multicolumn{1}{c}{$\delta(2000.0)$}&  
\multicolumn{1}{c}{V}& \multicolumn{1}{c}{B-V} &
\multicolumn{1}{c}{V-K$_{\rm s}$} &  \multicolumn{1}{c}{J-K$_{\rm s}$} &
\multicolumn{1}{c}{$RV_{\rm helio}$} & \multicolumn{1}{l}{S/N}\\
\multicolumn{1}{c}{}& \multicolumn{1}{c}{}&
\multicolumn{1}{c}{(hh mm s)}& \multicolumn{1}{c}{($\degr$ $\arcsec$ $\arcmin$)}&
\multicolumn{1}{c}{(mag.)}& \multicolumn{1}{c}{(mag.)} &
\multicolumn{1}{c}{(mag.)}& \multicolumn{1}{c}{(mag.)} & 
\multicolumn{1}{c}{(km s$^{-1}$)} & \multicolumn{1}{l}{at 6000 \AA } \\
\hline

NGC 2251 &   3 & 06:34:51.24 & +08:19:31.90 & 10.39 &$+$1.22 &$+$2.92 &$+$0.72 &
+26.2$\pm$0.2  & 150 \\
         &  33 & 06:34:37.07 & +08:21:39.50 & 10.39 &$+$1.21 &$+$2.91 &$+$0.70
&$+$26.3$\pm$0.2 & 150 \\
NGC 2266 &  73 & 06:43:16.69 & +26:57:05.19 & 11.06 &$+$0.99 &$+$2.46 &$+$0.69
&$-$29.7$\pm$0.2 & 100 \\
NGC 2335 &  11 & 07:06:11.42 &$-$09:56:22.70& 10.89 &$+$1.13 &$+$2.76 &$+$0.67
&$-$3.21$\pm$0.1 & 160 \\
NGC 2482 &   9 & 07:55:09.09 &$-$24:22:30.25 & 10.27 &$+$1.11 &$+$2.38 &$+$0.61
&+39.00$\pm$0.2 & 190 \\
NGC 2527 &  10 & 08:04:46.97 &$-$28:07:50.04 & 09.49 &$+$0.95 &$+$2.17 &$+$0.59
&+40.7$\pm$0.2 & 170 \\ 
         & 203 & 08:05:33.91 &$-$28:08:58.44 & 09.51 &$+$0.98 &$+$2.21 &$+$0.59
&+40.4$\pm$0.2 & 180 \\
NGC 2539 & 346 & 08:10:23.02 &$-$12:50:43.25 & 10.92 &$+$0.99 &$+$2.28 &$+$0.57
&+29.7$\pm$0.2 & 130 \\ 
         & 463 & 08:10:42.87 &$-$12:40:11.80 & 10.69 &$+$1.03 &$+$2.42 &$+$0.60
&+28.7$\pm$0.2 & 135 \\
NGC 2682 &  84 & 08:51:12.73 & +11:52:42.68  & 10.51 &$+$1.11 &$+$2.53 &$+$0.67
&+35.2$\pm$0.2 & 130 \\
         & 151 & 08:51:26.22 & +11:53:52.23  & 10.48 &$+$1.10 &$+$2.52 &$+$0.66
&+35.0$\pm$0.2 & 120 \\
         & 164 & 08:51:29.03 & +11:50:33.40  & 10.52 &$+$1.11 &$+$2.56 &$+$0.61
&+34.3$\pm$0.1 & 120 \\
    
\hline
\end{tabular}
\end{table*}

\section{Observations and data reduction}

Red giant members of OCs, with the exception of NGC 2682 (M67), not yet
 subjected to a comprehensive abundance analysis
using high resolution 
spectroscopy were selected from the \emph{New catalogue of optically visible open
clusters 
and candidates\footnote{http://www.astro.iag.usp.br/~wilton/}} (Dias et al. 2002). 
Selection of red giants instead of main sequence stars enables the extension
of our observations to more
distant OCs.
We have made use of the {\small WEBDA\footnote{http://www.univie.ac.at/webda/}}
database
for the selection of suitable red giant candidates.

Target clusters and their properties are shown in the Table 1: column 1
represents the cluster name, 
columns 2 \& 3 the Galactic longitude and latitude in degrees, 
column 4 the age, 
column 5 the photometric estimate of the iron abundance, column 6 the Galactocentric
distance, column 7 the distance modulus, 
column 8 the reddening, column 9 the reference to the photometric [Fe/H]. All
quantities
are from the database entry except for the photometric [Fe/H] abundance and 
the Galactocentric distance, R$_{\rm gc}$, which we calculate assuming 
a distance of the Sun from the Galactic centre of 8.0$\pm$0.6 kpc (Ghez et al.
2008).   

\begin{figure} 
\begin{flushleft} 
\includegraphics[width=8.8cm,height=9.5cm]{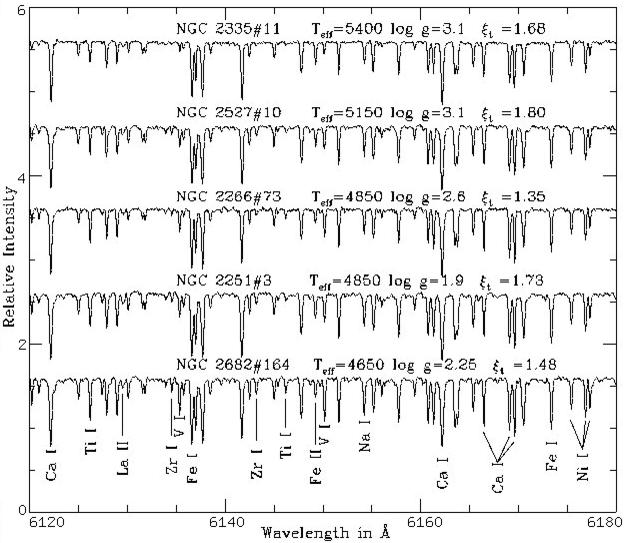}  \vskip-2ex
\caption{ Representative spectra of the red giant members of the five open clusters 
          described in Table 1}
 \label{spectra} 
\end{flushleft}
\end{figure}

High-resolution optical spectra of the program stars were obtained during the nights
of 
2011 January 12-14 and March 15-18 with the Robert G.
 Tull echelle coud\'{e} spectrograph (Tull et al. 1995)
on the 2.7-m Harlan J. Smith telescope at the McDonald observatory using a 2048 x
2048 pixel 
Tektronix charge-coupled device (CCD) as a detector. The spectra correspond to a
resolving power of
$\ga$ 55,000 ($<$ 6 km s$^{-1}$) as measured by the FWHM of Th {\scs I} lines
in 
comparison spectra. The spectral coverage in a single exposure from 4000 \AA\ to
5600 \AA\ 
across various orders was complete but incomplete
from 5600 \AA\ to about 9800 \AA, where the inter-order gaps begin to appear. 
In order to minimize the effect of cosmic rays and 
to acquire a good signal-to-noise (S/N) ratio for our 
spectra, two or three exposures were taken
each lasting for 20-30 min depending on the magnitude of the star.  

The basic observation procedure and the data reduction techniques are same as 
those listed in our paper {\scs I}.
The echelle spectroscopic data is extracted to one-dimensional spectra with y-axis
as flux and the x-axis 
as wavelength, using the standard spectral reduction software {\scs
IRAF}\footnote{IRAF is a general
purpose software system for the reduction and analysis of astronomical data
distributed by NOAO, 
which is operated by the  Association of Universities for Research in Astronomy, 
Inc. under cooperative agreement with the National Science Foundation.}.  
  
Multiple exposure frames from each of the stars were median combined to acquire
optimal S/N ratio 
and then trimmed to reduce the edge effects during the continuum fitting. The final
extracted spectra
have S/N ratio of about 100$-$190 as measured around 6000 \AA\ region, while at
wavelengths shorter than 
5000 \AA\ the S/N ratio decreases with decreasing wavelength.   
The continuum fitting was done interactively by marking continuum regions on an each
aperture which 
were then fitted by a cubic spline of appropriate order for proper continuum fitting. 

We measured the radial velocity (RV) of each star from its continuum-fitted 
spectrum using the cores of weak and 
moderately strong lines of various atomic species.
Observed radial velocities were converted to heliocentric velocities using the {\it
rvcorrect} routine available in {\scs IRAF} software. 
Our radial velocity measurements are in good agreement with 
the previous radial velocity measurements for the red giants in OCs (Mermilliod et
al. 2008). But, for NGC 2266, no radial velocity measurements from 
high-resolution spectra are available in the literature. 
Recently, Carrera (2012) analyzed 
medium-resolution spectra (R $\sim$ 8000) in the infrared CaT region ($\sim$8500 \AA)
for stars in NGC 2266 and derived a mean radial velocity of 
$\langle$ RV $\rangle$= $-$16$\pm$15 km s$^{-1}$ based on four stars. 
For this cluster, we  selected the star with ID 73 whose membership has been confirmed
through photometry (Kaluzny \& Mazur 1991). 
Our radial velocity estimate for this OC would appear to be the first
measurement available in the literature with negligible dispersion whereas 
Carrera's measurement suffers from large uncertainty around the mean.
Our radial velocity measurement for this OC differs from the Carrera's value 
by $-$10 km s$^{-1}$, but is well within the quoted uncertainty in Carrera (2012).

The identification and basic observational data for the stars observed in each of
the clusters are given in Table 2 along with the (V-K$_{\rm s}$) and (J-K$_{\rm s}$) 
colors from the Two Micron All Sky Survey (2MASS) 
catalogue\footnote{http://irsa.ipac.caltech.edu/applications/Gator} (Cutri et al.
2003)\footnote{Originally published in University of Massachusetts and 
Infrared Processing and Analysis Center (IPAC)/ California Institute of Technology.},  
computed radial velocity and S/N of each of the spectra extracted at 6000 \AA\ for
each of the stars. 
Spectra of a representative region are shown in Figure \ref{spectra}
for one star from a sample of the five OCs.

\section{Abundance analysis}
\subsection{\small Line selection}
Our selection criteria for suitable stellar features from each spectra 
followed the precepts discussed in paper {\scs I}. 
The equivalent widths (EWs) for each of the selected absorption features were
measured manually 
using the routine {\it splot} contained in {\scs IRAF} by fitting often a
Gaussian profile.
But for lines with significant damping wings a Voigt profile was used, for a few
lines a direct 
integration was preferred as a best measure of EW. 
The final linelist includes  250 lines of 23 elements
covering  the spectral range 4500 $\sim$ 8850 \AA. Our selection criteria 
provides, on average across the sample of 12 stars, a list of 60 Fe {\scs I}
lines with 
lower excitation potentials (LEPs) ranging from 0.9 to 5.0 eV, and 12 Fe
{\scs II} lines
with LEPs of 2.8 to 3.9 eV and EWs up to 120 m\AA.        
The line list is essentially that given in Table 3 of Paper {\sc i}.
  
Several lines may be affected by telluric absorption lines.
The EWs for these lines have been measured only when they appear to be Doppler
shifted away 
from the telluric components, as judged by referring  to  the
Arcturus spectrum (Hinkle et al. 2000).

\subsection{Determination of atmospheric parameters}

\subsubsection{Photometry}

The initial determination of effective temperature for each red giant was derived 
from dereddened\footnote{The adopted interstellar extinctions are  
(A$_{V}$, A$_{K}$, E(V-K), E(J-K))= (3.1, 0.28, 2.75, 0.54)*E(B-V), where
E(B-V) is taken from {\scs WEBDA} }
B, V, J and K photometry using the empirically calibrated
color-temperature relations by Alonso et al. (1999). The corresponding
errors around the (B-V), (V-K) and (J-K) relations are 167 K, 25 K and 125 K.  
Before that the 2MASS K$_{\rm s}$\footnote{The 2MASS survey uses a K-short 
(K$_{\rm s}$) filter whose 
effective wavelength is centred around 2.16 microns in the near-infrared. } 
magnitudes are transformed to standard K magnitude using the relations given in 
Carpenter (2001). The mean difference between the two magnitudes is 
K$_{\rm s}$=K+(-0.044$\pm$0.003). 

The surface gravities were computed by incorporating the known distance to 
the OCs, effective temperature $T_{\rm eff_\star}$, bolometric correction $BC_{V}$,
and 
the cluster turn-off mass $M_{\star}$ into the relation given by
(Allende Prieto et al. 1999).    

\noindent
 $\log{g_\star}$= $\log{g}_{\odot}$+$\log(M_{\star}/M_{\odot})$+4 $\log(T_{\rm
eff_\star}/T_{\rm eff}{\odot})$
 \begin{equation}
 \hspace{1.1cm}  +0.4(V_{0}+BC_{V})+2 log{\pi}+0.12  
\end{equation}
 with the corresponding luminosity given by
\begin{equation}
 log(L_{*}/L_{\odot})= -[0.4 (V_{0}+BC_{V})+2 log~{\pi}+0.12]  
 \end{equation}   \vskip1ex \noindent
 where $\pi$ is the parallax and V$_{0}$ is the dereddened Johnson V magnitude  
and the bolometric correction
 BC$_{V}$ are derived from estimated effective temperatures and metallicity
[Fe/H]$_{\rm phot.}$  
in Table \ref{tab1} from the calibration by Alonso et al. (1999). 
We adopt log~{g}$_{\odot}$= 4.44 cm s$^{-2}$ and T$_{\rm eff},_{\odot}$= 5777 K. 

A cluster's turn-off mass has been estimated from the stellar evolutionary tracks by
Yi et al. (2003):
turn-off masses are  3.3 $M_{\odot}$ for NGC 2251, 2.1 $M_{\odot}$ for NGC 2266,
3.8 $M_{\odot}$ for NGC 2335, 3.0 $M_{\odot}$ for NGC 2482, 2.8 $M_{\odot}$ for NGC
2527,
3.1 $M_{\odot}$ for NGC 2539 and 1.4 $M_{\odot}$ for NGC 2682.  
Red giants are assumed to have the turn-off mass in computing the surface gravity.

Assuming that the various quantities involved in equations (1) \& (2) are independent
of each other and by introducing an error of 10\% in the stellar mass, 
an uncertainty of 3\% in T$_{eff_\star}$, an uncertainty of 5\% in photometric V
magnitude
and the bolometric corrections, and an error of 10\% in the distance (i.e., the
parallax), we estimate an
error of $\simeq$ 0.11 dex in $\log~g_\star$ with an uncertainty of 0.08 in
$\log~L_{*}$.

\subsubsection{Spectroscopy}

The spectroscopic abundance analysis was executed using the latest version (2010) of
the local
thermodynamical equilibrium (LTE) line analysis and spectrum synthesis code
{\scs \bf MOOG} 
developed by Chris Sneden and originally described in 
Sneden (1973)\footnote{http://www.as.utexas.edu/~chris/moog.html}. Model atmospheres 
were interpolated linearly from the ATLAS9 model atmosphere grid of
Castelli \& Kurucz (2003)\footnote{http://kurucz.harvard.edu/grids.html}. 
For this purpose, initially we used a model with photometrically 
determined atmospheric parameters. These models assume a line-blanketed plane-parallel 
uniform atmospheres in LTE and hydrostatic equilibrium with flux conservation.

The linelist (Table 13 available online) and basic atomic line data (see  Table 3 of 
paper {\scs I}) 
give  reference
solar abundances that agree well with the published values from Asplund et al. (2009).
We performed a differential abundance analysis relative to the Sun
by running the {\scs \bf MOOG} in {\it abfind} mode using
the initially estimated stellar parameters (see., Section 3.2.1) and the measured EWs.
This {\it abfind} driver force-fits the individual line abundances to match
the computed EWs to the observed ones, previously 
measured using the {\it splot} routine of {\scs IRAF}. 
   
First, we estimated the 
microturbulence velocity, $\xi_{t}$, using Fe {\scs II} lines instead of Fe
{\scs I},
since Fe {\scs II} lines are less affected by the departures from LTE.
The  $\xi_{t}$ is determined by 
the requirement that the abundance from Fe {\scs II} line generally chosen to
have 
small range in LEP (2.8-3.9 eV) but a good range in their EWs, are independent of
line's EW or reduced
EW (log(EW/$\lambda$). The T$_{eff}$ is estimated by the requirement that the Fe
abundance from 
Fe {\scs I} lines is independent of a line's LEP. Finally, the surface gravity,
$\log~{g}$, is estimated from the constraint that Fe {\scs I} and
Fe{\scs II}
lines give the same Fe abundance for the derived T$_{\rm eff}$ and $\xi_{t}$.

\begin{figure} 
\begin{center} 
\includegraphics[width=8.3cm,height=8.2cm]{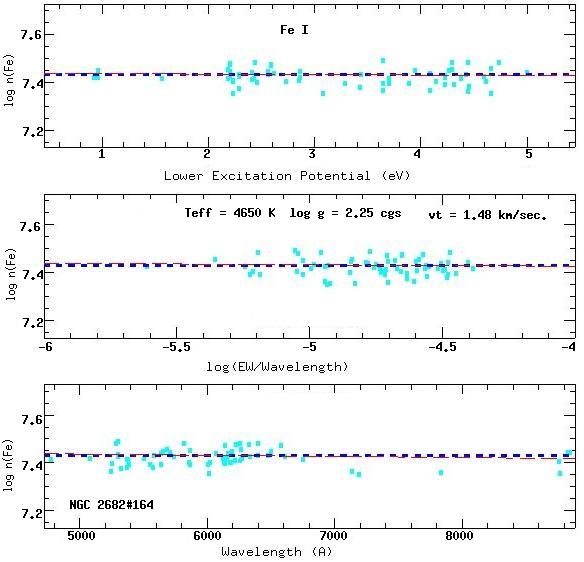} 
\caption{The observed trends for the derived abundance from the Fe I lines, [Fe I/H],
 for the star NGC 2682\#164 as a function of LEP (upper panel), reduced equivalent
width
 (lower panel), and the wavelength of each line(lower panel).}
 \label{moogplot}  
\end{center}
\end{figure}

\begin{table*}
\centering
 \begin{minipage}{165mm}
 \caption{Basic photometric and spectroscopic atmospheric parameters for the stars
in each cluster.}
 \label{tab3}
 \begin{tabular}{lcccccccccc}
  \hline
\multicolumn{1}{l}{Cluster}&  
\multicolumn{1}{c}{Star ID}&  
\multicolumn{3}{c}{T$^{\rm phot}_{\rm eff}$ (K)}&
\multicolumn{1}{c}{$\log g^{(V-K)}_{\rm phot}$}&  
\multicolumn{1}{c}{T$^{\rm spec}_{\rm eff}$}&  
\multicolumn{1}{c}{$\log g_{\rm spec}$}&
\multicolumn{1}{c}{$\xi_{\rm spec}$}&
\multicolumn{1}{c}{$\log(L/L_\odot)$} &
\multicolumn{1}{l}{$\log(L/L_\odot)$} \\  \cline{3-5} 
\multicolumn{1}{c}{}&
\multicolumn{1}{c}{}&
\multicolumn{1}{c}{(B-V)}& (V-K)& (J-K)&
\multicolumn{1}{c}{(cm s$^{-2}$)}&
\multicolumn{1}{c}{(K)}&
\multicolumn{1}{c}{(cm s$^{-2}$)}&
\multicolumn{1}{c}{(km sec$^{-1}$)}& \multicolumn{1}{l}{spectroscopy} &
\multicolumn{1}{l}{photometry} \\
\hline

NGC 2251 &  3 & 4620 & 4642 & 4575 & 2.21  & 4850 & 1.90 & 1.73 & 2.75 & 2.36 \\
         & 33 & 4693 & 4647 & 4624 & 2.24  & 4850 & 2.00 & 1.67 & 2.65 & 2.31 \\
NGC 2266 & 73 & 4953 & 4910 & 4564 & 1.80  & 4850 & 2.60 & 1.35 & 1.85 & 2.67 \\
NGC 2335 & 11 & 5372 & 5509 & 5145 & 2.51  & 5400 & 3.10 & 1.68 & 1.80 & 2.43 \\
NGC 2482 &  9 & 4802 & 4945 & 4756 & 2.34  & 4850 & 2.50 & 1.73 & 2.10 & 2.29 \\
NGC 2527 & 10 & 4946 & 4994 & 4740 & 2.80  & 5150 & 3.10 & 1.80 & 1.59 & 1.84 \\
         &203 & 4884 & 4950 & 4734 & 2.78  & 5050 & 2.80 & 1.61 & 1.85 & 1.84 \\
NGC 2539 &346 & 4934 & 5002 & 4888 & 2.63  & 5175 & 3.10 & 1.70 & 1.64 & 2.05 \\
         &463 & 4840 & 4844 & 4794 & 2.47  & 5050 & 2.80 & 1.68 & 1.90 & 2.15 \\
NGC 2682 & 84 & 4702 & 4683 & 4508 & 2.36  & 4800 & 2.60 & 1.55 & 1.66 & 1.85 \\
         &151 & 4720 & 4696 & 4552 & 2.36  & 4700 & 2.25 & 1.38 & 1.98 & 1.87 \\
         &164 & 4702 & 4658 & 4712 & 2.36  & 4650 & 2.25 & 1.48 & 1.96 & 1.85 \\

\hline
 \end{tabular}
 \end{minipage}
\end{table*}

In Figure \ref{moogplot}, we show a graphical representation of the determination of 
atmospheric parameters using Fe {\scs I} line abundances as a function of 
line's LEP, line strength, and the wavelength of the lines for a typical star
(NGC 2682\#164) together with a linear fit to each distribution. 
The vertical axis gives the logarithmic iron abundance on 
the standard scale in which log $\epsilon$(H) = 12. 
The thick dashed blue lines represents the mean Fe {\scs I} abundance, 
and the thin dashed red lines represent linear trends of abundance with the 
three variables. The lack of significant trends of the iron abundance with
LEP (upper panel) and line strength (middle panel) supports the validity of our 
chosen model and the adopted $\xi_{t}$. The absence of any trend between the derived 
iron abundance and the wavelength (bottom panel) represents a check on 
our continuum placement.  

A check on the derived spectroscopic atmospheric parameters was performed in the
same way as described in
paper {\scs I}. In short, the uncertainty in the microturbulence 
is estimated from the spread in 
the derived abundance as computed from Fe {\scs I}/Fe  {\scs II},
Ti {\scs I}/Ti {\scs II}, V {\scs I} and Cr {\scs I}/Cr
{\scs II},
around its minima while the microturbulence is varied over the range 
from 0 to 6 km s$^{-1}$ (See Figure 2 in our paper {\scs I} for illustration). The
uncertainty in the 
$\log~{g}$ value is provided by the ionization equilibrium between neutral and
ionized species of 
the elements Sc, Ti, V and Cr. 
Finally, the uncertainty in the T$_{\rm eff}$ is acquired by inspecting the 
slope of the relation between the Fe {\scs I} abundance and LEPs of the lines.
Therefore, the typical uncertainties estimated in this analysis are 100 K in
T$_{eff}$, 0.25 cm s$^{-2}$ 
in log~$g$ and 0.20 km s$^{-1}$ in $\xi_{t}$.

The derived stellar parameters for program stars in each of the cluster are shown in 
Table \ref{tab3}: column 1 represents the cluster name, column 2 represents the star
ID, 
columns 3 \& 4 represent the photometric T$_{\rm eff}$ using (B-V), (V-K) and (J-K)
and log~$g$ values, 
columns 5-7 represent the spectroscopic T$_{\rm eff}$, log~$g$ and $\xi_{t}$
estimates. 
Finally, the spectroscopic and photometric luminosities ( log(L/L$_{\odot})$)
are presented in columns 8 \& 9.
 The mean difference in effective temperatures
estimated using (B-V) and (V-K) is $-$26 $\pm$ 64 K and using (V-K) and (J-K) is
$+$157 $\pm$ 121 K.
The corresponding mean differences in T$^{(B-V)}_{\rm eff}$, 
T$^{(V-K)}_{\rm eff}$ and T$^{(J-K)}_{\rm eff}$ with respect to 
spectroscopic T$^{spec}_{\rm eff}$'s are $-$100 $\pm$ 113K, $-$75 $\pm$ 116K and
$-$232 $\pm$ 117K respectively.
Mean differences in log~$g$ and log(L/L$_{\odot}$) across the 12 stars are
$-0.18\pm0.33$ cm-s$^{-2}$
and $-0.15\pm0.35$ cm-s$^{-2}$ respectively.

The photometric T$_{\rm eff}$'s derived from the calibrations based on
infrared flux method (Alonso et al. 1999)   
are sensitive to the adopted colors, reddening estimates and 
metallicity. 
When using the (B-V)$-$ T$_{\rm eff}$ calibration an error of 0.03 mag on (B-V)
and a conservative uncertainty of 20 \% in reddening translates into a temperature 
uncertainty of 1.1 to 1.3 \% each.  
Equivalently, an error of 0.2 dex in [Fe/H]$_{\rm phot}$ 
implies a temperature uncertainty of 1.2 \%. 
When using the (V-K)$-$ T$_{\rm eff}$ calibration, we note that an error of 0.03 mag
on (V-K)
and an uncertainty of 20 \% in reddening implies a temperature uncertainty of 0.1 to
0.7 \% each and 
an error of 0.2 dex in [Fe/H]$_{\rm phot}$ implies a temperature uncertainty of 0.1
to 0.9 \%.
Even though the T$_{\rm eff}$ as a function of (J-K) has no dependence on
metallicity, an error of 
0.03 mag on (J-K) implies a temperature uncertainty of 2 \% while the effect of
reddening 
variation is less.
As we have shown in Paper {\sc i}, temperatures derived from (V-K) colour might be
least affected by photometric
uncertainties and we shall use these values as the photometric T$_{\rm eff}$
estimates of our giant stars. 

\begin{figure*} 
\begin{center} 
\includegraphics[width=17cm,height=6.5cm]{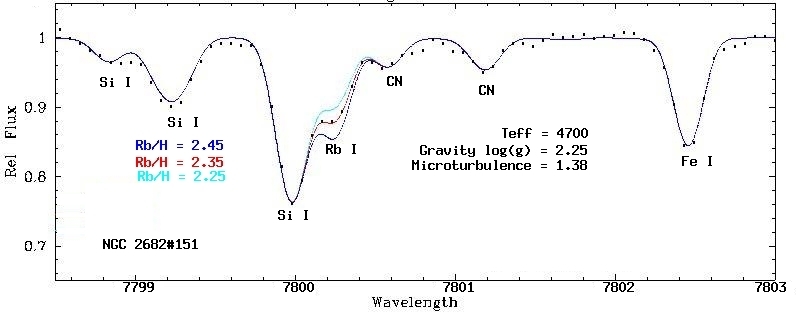}  \vskip-2ex
\caption{Synthetic spectra (continuous lines) and the observed spectrum (dotted
line) of NGC 2682 \#151 around the
  Rb I line at 7800 \AA. The indicated abundances in the figure are on a logarithmic
scale.}
 \label{synthrb} 
\end{center}
\end{figure*}

With few  exceptions, our spectroscopic estimates are in good agreement with 
the photometric ones. The uncertainty affecting  latter is subject to
the uncertainties
in the reddening values. For example, for  NGC 2251 the difference in
T$^{(V-K)}_{\rm eff}$
and T$^{\rm spec}_{\rm eff}$ is around 200 K using the reddening estimate of E(B-V)=
0.186 
as quoted in {\small WEBDA} database. If we replace this value with the one (E(B-V)=
0.21$\pm$0.03)
given in the recent analysis of Parisi et al. (2005), the difference between the two
temperatures
will be reduced by 50-100 K. 
The corresponding comparison of the spectroscopic [Fe/H] with the photometric one in
Table 1 also illustrates fair agreement: $\Delta$[Fe/H] = 0.10 (NGC 2251),
 $-$0.05 (NGC 2266), $-$0.15 (NGC 2335),
 $-$0.19 (NGC 2482), $-$0.01 (NGC 2527), 0.14 (NGC 2539) and $-$0.08 (NGC 2682). 

\subsection{Abundances and error estimation}

Once the atmospheric parameters were estimated from the spectral line measurements
as described in previous section, the corresponding model atmospheres were used 
in {\it abfind} and {\it synth}  drives of {\scs \bf MOOG} to conduct a 
complete abundance analysis.
In most cases, the abundances are derived from the measured EWs but a few
lines were analysed with synthetic spectra. 

Synthetic spectra  were computed for
species affected by hyperfine and isotopic splitting or affected by
blends. Llines treated by synthesis included:
Sc {\scs II} (6245 \AA), 
Mn {\scs I} (6013 \AA\ and 6021 \AA), V {\scs I} (5727 \AA), 
Cu {\scs I} (5218 \AA), Rb {\scs I} (7800 \AA),  
Ba {\scs II} (5853 \AA) and Eu {\scs II} (6645 \AA)
by including their hfs components, and Ce {\scs II} (5472 \AA) and 
Sm {\scs II} (4577 \AA).
We  adopted the hfs data and isotopic ratios listed  in
Paper {\scs I}. The hfs data and isotopic ratios for the 
Eu {\scs II} line has been kindly provided by C. Sneden whose linelist uses the 
$gf$-values provided by Lawler et al. (2001). 
For the Rb {\scs I} line,
the number of hfs components, the accurately known wavelengths and the relative 
strength of each component were taken from Lambert \& Luck's (1976) analysis of 
the Rb {\scs I} resonance line in the Arcturus spectrum and a reliable 
$\log~{gf}$ value of +0.13$\pm$ 0.04 from Wiese \& Martin (1980). A fit to this line 
includes a Si {\scs I} blend to the blue. As there are no reliable measurements
of experimental $gf$-values for this line, we have adopted a solar $gf$-value 
of $-$0.75 to model the solar Si {\scs I} line profile from  
the solar integrated disc spectrum (Kurucz et al. 1984). 
Our solar abundance measurement of $\log~\epsilon_{\odot}$ = 2.60 dex 
for the Rb is in close agreement with Asplund et al's. (2009)
value of $\log~\epsilon_{\odot}$ = 2.52$\pm$0.10 dex. 

A fit to the Rb {\scs I} line at 7800 \AA\ in the observed spectrum (dotted line) 
of NGC 2682 \#151
is shown in figure \ref{synthrb} as solid lines with [Rb/H] = 2.35 $\pm$ 0.10 dex,
based on $\chi^{2}$ goodness of fit provided by MOOG.   
A visual inspection of the figure shows that the red line ([Rb/H] = 2.35) is a 
better match to the stellar spectrum.

The largest hfs correction is about $-$0.11 dex for [Sc/Fe], $-$0.10 dex for [V/Fe],
$-$0.43 dex for [Mn/Fe], $-$0.39 dex for [Cu/Fe], $-$0.12 dex for [Ba/Fe], 
$-$0.32 dex for [Eu/Fe]. 

Abundance results for the individual stars in each of the OCs are presented in 
Tables \ref{NGC2527}$-$\ref{NGC2482}.
For each abundance based on the analysis of EWs, the abundance and standard
deviation were 
calculated from all lines of given species. The tables give the [Fe/H] and [X/Fe] for 
elements considered here relative to solar abundances derived from the adopted
$gf$-values. Therefore, errors in the adopted gf values are unimportant when providing 
differential abundances ([X/H] or [X/Fe]) provided that the solar and stellar
abundances depend on the same set of lines.

\begin{table}
\centering
 \begin{minipage}{90mm}
 \caption{Sensitivity of abundances to the uncertainties in the model
 parameters for the star with ID 11 in NGC 2335 with $T_{\rm eff}$= 5400 K,
$\log{g}$= 3.10
cm s$^{-2}$,and $\xi_{t}$= 1.68 km s$^{-1}$.}
 \label{sensitivity}
 \begin{tabular}{lllll}
  \hline
\multicolumn{1}{l}{ }&\multicolumn{1}{l}{$T_{\rm eff}\pm$100
K}&\multicolumn{1}{l}{$\log~{g}\pm$ 0.25} &\multicolumn{1}{l}{$\xi_{t}\pm$ 0.20} &
\multicolumn{1}{l}{ } \\ \cline{2-5}
\multicolumn{1}{l}{Species}&\multicolumn{1}{c}{$\sigma_{T_{eff}}$}&\multicolumn{1}{c}{$\sigma_{log{g}}$}
 & \multicolumn{1}{c}{$\sigma_{\xi_{t}}$} &\multicolumn{1}{c}{$\sigma_{2}$}  \\
\hline

Na {\scs I} & $+0.05/-0.06$ &$-0.02/+0.02$  &$-0.04/+0.05$   & 0.04   \\
Mg {\scs I} & $+0.03/-0.03$ &$-0.01/+0.01$  &$-0.02/+0.03$   & 0.02   \\
Al {\scs I} & $+0.03/-0.04$ &$-0.01/+0.01$  &$-0.02/+0.02$   & 0.02   \\
Si {\scs I} & $+0.01/-0.01$ &$+0.03/-0.02$  &$-0.02/+0.02$   & 0.02   \\
Ca {\scs I} & $+0.07/-0.08$ &$-0.02/+0.02$  &$-0.09/+0.09$   & 0.07   \\
Sc {\scs I} & $+0.09/-0.09$ &$~~0.00/+0.01$ &$~~0.00/+0.01$   & 0.05   \\
Sc {\scs II}& $-0.02/+0.01$ &$+0.12/-0.11$  &$-0.05/+0.06$   & 0.07   \\ 
Ti {\scs I} & $+0.10/-0.12$ &$-0.01/~~0.00$  &$-0.03/+0.03$   & 0.06   \\
Ti {\scs II}& $-0.03/+0.01$ &$+0.11/-0.12$  &$-0.07/+0.08$   & 0.08   \\
V {\scs I}  & $+0.12/-0.13$ &$-0.01/~~0.00$  &$-0.01/+0.01$   & 0.05   \\
Cr {\scs I} & $+0.08/-0.10$ &$-0.01/~~0.00$  &$-0.07/+0.07$   & 0.06   \\ 
Cr {\scs II}& $-0.05/+0.04$ &$+0.11/-0.11$  &$-0.05/+0.06$   & 0.07   \\ 
Mn {\scs I} & $+0.07/-0.09$ &$~~0.00/-0.01$  &$-0.04/+0.04$   & 0.05   \\
Fe {\scs I} & $+0.08/-0.09$ &$~~0.00/~~0.00$  &$-0.09/+0.10$   & 0.07   \\
Fe {\scs II}& $-0.07/+0.05$ &$+0.12/-0.13$  &$-0.07/+0.07$   & 0.09   \\ 
Co {\scs I} & $+0.08/-0.08$ &$+0.01/-0.01$  &$-0.01/+0.02$   & 0.05   \\ 
Ni {\scs I} & $+0.07/-0.06$ &$+0.02/-0.02$  &$-0.05/+0.05$   & 0.05   \\ 
Cu {\scs I} & $+0.07/-0.08$ &$+0.02/-0.02$  &$-0.04/+0.06$   & 0.05   \\
Zn {\scs I} & $+0.02/+0.03$ &$+0.07/-0.06$  &$-0.05/+0.07$   & 0.05   \\
Rb {\scs I} & $+0.07/-0.08$ &$-0.01/~~0.00$  &$-0.01/~~0.00$   & 0.04   \\
Y {\scs II} & $-0.01/-0.01$ &$+0.11/-0.12$  &$-0.07/+0.09$   & 0.08   \\
Zr {\scs I} & $+0.13/-0.13$ &$~~0.00/-0.01$  &$~~0.00/+0.01$   & 0.07   \\
Zr {\scs II}& $-0.01/~~0.00$ &$+0.10/-0.11$  &$-0.01/+0.01$   & 0.06   \\
Ba {\scs II}& $+0.01/-0.03$ &$+0.08/-0.10$  &$-0.17/+0.17$   & 0.11   \\
La {\scs II}& $~~0.00/-0.02$ &$+0.10/-0.12$  &$-0.02/+0.01$   & 0.06   \\
Ce {\scs II}& $-0.01/-0.01$ &$+0.10/-0.11$  &$-0.01/+0.01$   & 0.06   \\
Nd {\scs II}& $+0.01/-0.02$ &$+0.11/-0.11$  &$-0.02/+0.02$   & 0.06   \\
Sm {\scs II}& $+0.02/-0.01$ &$+0.11/-0.11$  &$-0.02/+0.03$   & 0.06   \\
Eu {\scs II}& $-0.01/~~0.00$ &$+0.11/-0.11$  &$-0.01/+0.01$   & 0.06   \\

\hline
 \end{tabular}
 \end{minipage}
\end{table}

Inspection of the Tables \ref{NGC2527}$-$\ref{NGC2482} shows that, in general, the
compositions [X/Fe] of stars in a given cluster are generally identical to within the
(similar) standard deviations computed for an individual star.  Exceptions tend to
occur for species represented by one or a few lines, as expected when the
uncertainty in
measuring equivalent widths is a contributor to the total uncertainty.  
From the spread in the abundances for the stars of a given cluster we obtain
the standard deviation $\sigma_1$ in the Tables \ref{NGC2527}$-$\ref{NGC2251} in the 
column headed `average'. 

\begin{table*}
 \begin{minipage}{160mm}
 \caption{Elemental abundance ratios [X/Fe] for elements from Na to Eu for a sample
seven OCs from this study }
 \label{abundance}
 \begin{tabular}{lllllccc}
  \hline
\multicolumn{1}{l}{Species}  & \multicolumn{1}{l}{NGC 2527} & \multicolumn{1}{l}{NGC
2682} &  
\multicolumn{1}{l}{NGC 2482} & \multicolumn{1}{l}{NGC 2539} & \multicolumn{1}{c}{NGC
2335} & 
\multicolumn{1}{l}{NGC 2251} & \multicolumn{1}{l}{NGC 2266}  \\
\hline

$[$Na {\scs I}/Fe$]$  &$+0.32\pm0.03$ &$+0.25\pm0.03$ &$+0.30\pm0.03$ 
&$+0.27\pm0.03$ &$+0.24\pm0.03$ &$+0.33\pm0.04$ &$+0.23\pm0.03$ \\
$[$Mg {\scs I}/Fe$]$  &$+0.07\pm0.01$ &$+0.16\pm0.02$ &$+0.13\pm0.02$ 
&$+0.07\pm0.02$ &$+0.08\pm0.02$ &$+0.06\pm0.02$ &$+0.39\pm0.02$ \\
$[$Al {\scs I}/Fe$]$  &$+0.05\pm0.02$ &$+0.09\pm0.01$ &$+0.07\pm0.02$  &$
0.00\pm0.01$ &$-0.02\pm0.02$ &$ 0.00\pm0.02$ &$+0.25\pm0.02$ \\
$[$Si {\scs I}/Fe$]$  &$+0.20\pm0.02$ &$+0.20\pm0.02$ &$+0.23\pm0.04$ 
&$+0.17\pm0.02$ &$+0.10\pm0.02$ &$+0.23\pm0.02$ &$+0.28\pm0.02$ \\
$[$Ca {\scs I}/Fe$]$  &$+0.12\pm0.04$ &$+0.04\pm0.04$ &$+0.01\pm0.05$ 
&$+0.04\pm0.04$ &$+0.09\pm0.04$ &$+0.09\pm0.05$ &$+0.17\pm0.05$ \\
$[$Sc {\scs I}/Fe$]$  &$+0.25\pm0.04$ &$+0.04\pm0.04$ &$+0.12\pm0.05$ 
&$+0.21\pm0.04$ &$+0.13\pm0.04$ &$+0.04\pm0.04$ &$+0.31\pm0.05$ \\
$[$Sc {\scs II}/Fe$]$ &$+0.12\pm0.04$ &$+0.10\pm0.04$ &$+0.08\pm0.05$ 
&$+0.04\pm0.04$ &$\bf+0.16$     &  $\bf+0.02$ &$+0.22\pm0.05$ \\ 
$[$Ti {\scs I}/Fe$]$  &$+0.11\pm0.04$ &$-0.01\pm0.04$ &$+0.01\pm0.04$ 
&$+0.10\pm0.04$ &$+0.17\pm0.04$ &$-0.04\pm0.04$ &$+0.23\pm0.04$ \\
$[$Ti {\scs II}/Fe$]$ &$+0.08\pm0.04$ &$+0.01\pm0.05$ &$-0.03\pm0.04$ 
&$+0.08\pm0.04$ &$+0.12\pm0.04$ &$-0.04\pm0.04$ &$+0.30\pm0.04$ \\
$[$V {\scs I}/Fe$]$   &$+0.21\pm0.04$ &$+0.09\pm0.05$ &$+0.10\pm0.05$ 
&$+0.16\pm0.04$ &$+0.13\pm0.03$ &$-0.05\pm0.05$ &$+0.20\pm0.06$ \\
$[$Cr {\scs I}/Fe$]$  &$+0.10\pm0.03$ &$+0.05\pm0.03$ &$+0.10\pm0.04$ 
&$+0.08\pm0.04$ &$+0.07\pm0.03$ &$+0.05\pm0.04$ &$+0.09\pm0.03$ \\ 
$[$Cr {\scs II}/Fe$]$ &$+0.06\pm0.04$ &$+0.08\pm0.05$ &$+0.09\pm0.05$ 
&$+0.08\pm0.04$ &$ 0.00\pm0.04$ &$+0.04\pm0.05$ &$+0.02\pm0.04$ \\ 
$[$Mn {\scs I}/Fe$]$  &  $\bf 0.00 $  &   $\bf-0.08$  &   $\bf-0.11$   &$\bf+0.01$  
  & $\bf-0.01$    &$\bf-0.13$     &$\bf-0.01$     \\
$[$Fe {\scs I}/H$]$   &$-0.11\pm0.04$ &$-0.08\pm0.04$ &$-0.07\pm0.04$ 
&$-0.06\pm0.04$ &$-0.19\pm$0.04 &$-0.10\pm0.05$ &$-0.45\pm$0.04 \\
$[$Fe {\scs II}/H$]$  &$-0.09\pm0.05$ &$-0.08\pm0.05$ &$-0.07\pm0.05$ 
&$-0.07\pm0.05$ &$-0.17\pm$0.05 &$-0.10\pm0.05$ &$-0.43\pm$0.05 \\ 
$[$Co {\scs I}/Fe$]$  &$+0.16\pm0.04$ &$+0.11\pm0.02$ &$+0.11\pm0.03$ 
&$+0.07\pm0.02$ &$+0.16\pm$0.04 &$+0.03\pm0.03$ &$+0.27\pm0.03$ \\ 
$[$Ni {\scs I}/Fe$]$  &$+0.06\pm0.02$ &$+0.10\pm0.03$ &$+0.03\pm0.04$ 
&$+0.02\pm0.02$ &$+0.09\pm$0.03 &$+0.04\pm0.03$ &$+0.09\pm0.03$ \\ 
$[$Cu {\scs I}/Fe$]$  &  $\bf-0.14$   &  $\bf-0.03$   &  $\bf-0.21$    & $\bf-0.16$ 
  &$\bf-0.15$     &$\bf-0.22$     &$\bf+0.02$     \\
$[$Zn {\scs I}/Fe$]$  &  $\bf-0.16$   &  $\bf-0.07$   &  $\bf-0.22$    & $\bf-0.22$ 
  &$\bf-0.06$     &$\bf-0.13$     &$\bf 0.00$     \\  
$[$Rb {\scs I}/Fe$]$  &  $\bf+0.07$   &  $\bf-0.10$   &  $\bf-0.13$    & $\bf+0.04$ 
  &$\bf+0.08$     &$\bf-0.17$     &$\bf+0.14$     \\ 
$[$Y {\scs II}/Fe$]$  &$+0.16\pm0.04$ &$+0.03\pm0.04$ &$+0.15\pm0.04$ 
&$+0.17\pm0.04$ &$+0.12\pm$0.05 &$+0.07\pm0.04$ &$+0.05\pm0.04$ \\
$[$Zr {\scs I}/Fe$]$  &$+0.31\pm0.05$ &$-0.07\pm0.05$ &$+0.11\pm0.05$ 
&$+0.21\pm0.04$ &$+0.06\pm$0.04 &$ 0.00\pm0.05$ &$-0.08\pm$0.05 \\
$[$Zr {\scs II}/Fe$]$ &$+0.18\pm0.03$ &$-0.07\pm0.03$ &$+0.10\pm0.04$  &$+0.26\pm0.03$ 
&$+0.01\pm$0.03 &$+0.06\pm0.03$ &$\ldots$      \\
$[$Ba {\scs II}/Fe$]$ & $\bf+0.08$    &   $\bf-0.16$  &   $\bf+0.09$   &$\bf+0.10$  
  &$\bf+0.25$     &$\bf+0.11$     &$\bf-0.13$     \\
$[$La {\scs II}/Fe$]$ &$+0.26\pm0.03$ &$ 0.00\pm0.03$ &$+0.18\pm0.03$ 
&$+0.18\pm0.03$ &$+0.29\pm$0.03 &$+0.02\pm0.04$ &$-0.02\pm0.04$ \\
$[$Ce {\scs II}/Fe$]$ &$\bf+0.24$     &$\bf-0.02$     &$\bf+0.11 $     &$\bf+0.20$  
  &$\bf+0.29$     &$\bf+0.07$     &$\bf-0.07$     \\  
$[$Nd {\scs II}/Fe$]$ &$+0.20\pm0.03$ &$+0.02\pm0.04$ &$+0.13\pm0.04$ 
&$+0.23\pm0.04$ &$+0.32\pm$0.03 &$+0.08\pm0.04$ &$+0.13\pm0.04$ \\
$[$Sm {\scs II}/Fe$]$ &$\bf+0.18$     &$\bf-0.03 $    &$+0.13\pm0.04$  &$\bf+0.18$  
  &$+0.28\pm$0.04 &$\bf+0.02$     &$\bf+0.11$     \\
$[$Eu {\scs II}/Fe$]$ &$\bf+0.10$     & $\bf+0.08$    &$\bf+0.07$      &$\bf+0.19$  
  &$\bf+0.07$     &$\bf+0.04$     &$\bf+0.40$     \\

\hline
\end{tabular}
 \flushbottom{{\bf Note}: Abundances calculated by synthesis are presented in bold
numbers.}
\end{minipage}
\end{table*}

We evaluated the sensitivity of the derived abundances to the uncertainties
in the adopted atmospheric parameters by varying each time 
only one of the parameters by the amount corresponding to the typical error. 
The changes in abundances caused by varying atmospheric parameters by 100 K, 
0.25 cm s$^{-2}$ and 0.2 km s$^{-1}$ with respect to the chosen model atmosphere are
summarized in Table \ref{sensitivity}. 
We quadratically added the three contributors, by taking the square root of the sum
of the square of individual errors associated with uncertainties in temperature,
gravity and microturbulence, 
to obtain $\sigma_{2}$. The total error $\sigma_{tot}$ for each of the element is
the quadratic sum of $\sigma_{1}$ and $\sigma_{2}$. The error bars in the abundance
tables correspond to this total error. The final OC mean abundances from this study
are presented in Table \ref{abundance}.

\begin{table}

 \begin{minipage}{83mm}
 \caption{ Mean elemental abundance ratios, [X/Fe], for Na to Eu for the six
 OCs from this study and the thin disc mean abundances from Luck \& Heiter (2007)
 in the metallicity of our clusters (0.0 to $-$0.2 dex). Abundances calculated by 
 synthesis are presented in bold typeface.}
\centering
 \label{abundance1}
 \begin{tabular}{lcc}
  \hline
 \multicolumn{1}{c}{Species}  & \multicolumn{1}{c}{OC mean} & \multicolumn{1}{c}{Thin disc}  \\
\hline

$[$Na {\scs I}/Fe$]$  &$+0.28\pm0.04$ & $+0.10\pm0.06$ \\
$[$Mg {\scs I}/Fe$]$  &$+0.09\pm0.04$ & $+0.08\pm0.10$ \\
$[$Al {\scs I}/Fe$]$  &$+0.03\pm0.04$ & $+0.09\pm0.05$ \\
$[$Si {\scs I}/Fe$]$  &$+0.19\pm0.05$ & $+0.12\pm0.04$ \\
$[$Ca {\scs I}/Fe$]$  &$+0.07\pm0.04$ & $-0.04\pm0.05$ \\
$[$Sc {\scs I}/Fe$]$  &$+0.13\pm0.09$ & $-0.08\pm0.06$ \\
$[$Ti {\scs I}/Fe$]$  &$+0.06\pm0.08$ & $ 0.00\pm0.03$ \\
$[$V {\scs I}/Fe$]$   &$+0.11\pm0.09$ & $-0.09\pm0.07$ \\
$[$Cr {\scs I}/Fe$]$  &$+0.07\pm0.02$ & $+0.01\pm0.05$ \\ 
$[$Mn {\scs I}/Fe$]$  &$\bf-0.05\pm0.06$ & $+0.06\pm0.07$ \\
$[$Co {\scs I}/Fe$]$  &$+0.11\pm0.05$ & $+0.06\pm0.08$ \\ 
$[$Ni {\scs I}/Fe$]$  &$+0.06\pm0.03$ & $ 0.00\pm0.03$ \\ 
$[$Cu {\scs I}/Fe$]$  &$\bf-0.15\pm0.07$ & $+0.01\pm0.13$ \\
$[$Rb {\scs I}/Fe$]$  &$\bf-0.04\pm0.11$ & $\ldots$ \\
$[$Y {\scs II}/Fe$]$  &$+0.12\pm0.06$ & $+0.07\pm0.15$ \\
$[$Zr {\scs I}/Fe$]$  &$+0.10\pm0.14$ & $\ldots$       \\
$[$Zr {\scs II}/Fe$]$ &$+0.09\pm0.12$ & $\ldots$       \\
$[$Ba {\scs II}/Fe$]$ & $\bf+0.09\pm0.15$ & $+0.04\pm0.16$ \\
$[$La {\scs II}/Fe$]$ &$+0.15\pm0.12$ & $+0.05\pm0.09$ \\
$[$Ce {\scs II}/Fe$]$ &$\bf+0.15\pm0.12$ & $+0.05\pm0.09$ \\  
$[$Nd {\scs II}/Fe$]$ &$+0.16\pm0.11$ & $-0.01\pm0.07$ \\
$[$Sm {\scs II}/Fe$]$ &$+0.13\pm0.11$ & $\ldots$ \\
$[$Eu {\scs II}/Fe$]$ &$\bf+0.09\pm0.05$ & $+0.08\pm0.06$ \\

\hline
\end{tabular} 
\end{minipage}
\end{table}

\section{Results}

The addition of these seven clusters to literature sample also supports the
widely held 
impression that there is an abundance gradient such that the metallicity [Fe/H] at 
the solar galactocentric distance decreases outwards (Magrini et al. 2009, Yong et
al. 2012)
at about $-$0.1 dex kpc$^{-1}$ and at the same time our 
sample has not changed the dispersion of metallicities at a given Rgc.

As in Paper {\scs I}, we compare our results with those from Luck \& Heiter (2007)
for a large sample of nearby giants whose analysis was similar to our, i.e., a
differential analysis with respect to the Sun. 
Perhaps, the chief advantage of this comparison is that various systematic 
effects should cancel which would enter into  a comparison involving our giants
and field main sequence stars. Two obvious issues are non-LTE effects which 
are, in general, going to be different for giants and dwarfs and the use of
classical atmospheres which simulations of stellar granulation show may
in different ways be inadequate representation of real stars.

From Luck \& Heiter's  Table 4, we calculate mean abundances [X/Fe] for [Fe/H] from
0.0 to $-$0.2,
the metallicity spread of six of our seven clusters. NGC 2266 at [Fe/H] = $-$0.44 is
excluded from the sample in part because it may belong to the thick rather than the
thin disc. One can argue based on the membership probabilities 
(see for example Reddy et al. 2006 for definitions and recipes) derived using the 
present day cluster's space motions that this cluster has a high probability to 
belong to the thick disc (Reddy \& Giridhar, in preparation).
Inspection of Table \ref{abundance1}
shows that the mean [X/Fe]  for the six clusters are within 0.10 dex of Luck \&
Heiter's results except for Na, Ca, Sc, V, Mn, Cu, and Nd and
within 0.15 dex for all but Na, Sc, V, and Nd. (Elements Zn, Rb, and Sm were not
considered by Luck \& Heiter.) Thus, we conclude that field giants and OCs of
near-solar metallicity have very similar, if not identical, compositions.
This was the conclusion reached also in Paper {\sc i}.

We review our conclusions made in our paper {\scs I} for Mn {\scs I}. 
Our [Mn/Fe] ratios for OC giants matches well with Takeda (2008) sample of 
field giants, while it is lower by 0.12 dex with respect to 
the local thin disc field giants of Luck \& Heiter (2007). 
Luck \& Heiter's (2007) abundance results for field giants are generally
confirmed by Takeda et al. (2008) for other large sample of field giants.
Indeed, Luck \& Heiter's (2007) [Mn/Fe] abundance ratios for red giants are 0.1 dex
lower than their 
sample of field dwarfs (Luck \& Heiter 2006). Both Luck \& Heiter's (2007) and
Takeda et al. (2008)
have employed similar abundance analysis based on the EW measurement of a large set
of lines.
Determination of the abundances of the elements using EW measurements of the lines
is not always
recommended as a good approximation for the features affected by hfs components.
Moreover, these
lines are typically strong in red giant branch stars than their main sequence
counterparts with 
similar metallicities.

Since, the hfs effect desaturates strong absorption lines and results in features
with larger 
equivalent widths the computed abundances will be overestimated. It will be more
complicated 
for the elements having isotopes due to the superposition of the spectra of the
various isotopes 
of the same element. The relative strength and isotopic shift vary from element to
element. 
Since, the stellar surface abundance estimates are higly dependent on the line
strength,
it is important to treat strong lines for hfs effects since the hfs correction 
increases with line strength.
For weak lines, those generally lay on the linear part of the curve of growth, 
the hfs treatment is not so important because they are already unsaturated. 
Hence a spectrum synthesis over the separate hfs components is recommended
to derive an abundance for the lines affected by hfs components.
Thus, the number of components, the wavelength splittings, isotopic shifts, if any
present,
and the relative strength of each the component must be known. 

In our analysis largest hfs corrections are applied to [Mn/Fe] ($\simeq$ $-$0.43 dex). 
From the Figure 15 of Takeda (2007) for a sample 
field dwarfs, it is evident that the Mn {\scs I} lines employed for their abundance
analysis
are all weak (EW $<$ 50 m\AA) for which they have noted a negligible hfs corrections
($\simeq$ $-$0.02 dex). Similarly, in their giant sample Takeda et al.(2008) have
employed a single weak line at 5004.8 \AA\ for the abundance analysis for 
which the hfs effects are less important. 
Whereas the Mn {\scs I} abundances from Luck \& Heiter (2007), similar to ours,
might be based on strong lines for which they have not employed hfs treatment,
unlike ours. 
This can fairly explain the observed offset in the abundance trends of field dwarfs 
and our OCs red giants with the abundance trend of field giants from Luck \& Heiter
(2007).

\begin{figure} 
\begin{flushleft} 
\includegraphics[width=7.8cm,height=5.0cm]{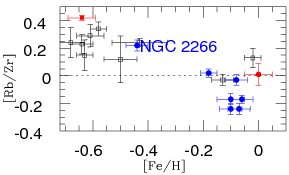}  \vskip-2ex
\caption{ The abundance ratio for [Rb/Zr] as a function of [Fe/H] with corresponding
error bars. 
 The dots (blue) represent the mean cluster
 abundances derived in this study. The OC NGC 2266 with a largest [Rb/Zr] ratio (=
+0.22) is marked 
 in the figure. The empty squares (black) represents the data taken
 from Tomkin \& Lambert (1999) for a sample of metal-deficient disc and halo stars.
The filled square 
 and the filled circle (in red) represents the [Rb/Zr] ratios for Arcturus and Sun. 
 The dotted line indicate the solar mix of elements.}
 \label{rbzr} 
\end{flushleft}
\end{figure}

The Rb abundance provides an additional diagnostic information on the neutron
density at the s-process
site which is controlled by both the neutron source and the mass of the parent AGB
star responsible 
for its synthesis. The abundance of Rb relative to its neighboring s-process
elements such as 
Sr, Y and Zr is a useful monitor of neutron density at the time of s-processing.
Thanks to the 
branch in the s-process path at $^{85}$Kr with a $\beta$-decay half life of 10.7 yr,
which at low 
neutron densities (N$_{n}$ $\leq$ 10$^7$ cm$^{-3}$) decays to $^{85}$Rb, while at
higher neutron
densities suffers successive neutron captures and the flow proceeds along the path 
from $^{85}$Kr to $^{86}$Kr to $^{87}$Kr, that $\beta$-decays to stable $^{87}$Rb
isotope.
The sensitivity of the Rb elemental abundance to the neutron density arises then due
to the 
quite different neutron capture cross sections for both of its stable isotopes 
($^{85}$Rb and $^{87}$Rb) among which the $^{85}$Rb being roughly a factor 10 times
more efficient
than $^{87}$Rb. Thus, at low neutron densities the ratio [Rb/Zr] is about 10 times
less 
than that at high neutron densities. Since, the solar Rb abundance is attributed in
almost
equal parts to the s-process and r-process, the observed stellar Rb abundance
directly indicate the 
neutron density at the s-process site. Even though the stellar Rb {\scs I} isotopic
ratios can not 
be measured, the Rb {\scs I} lines used in our analysis are weak in all the program
stars 
so the derived Rb abundances are not affected by changes in the isotopic mixture. We
derived the 
Zr abundances using 5-7 weak lines (see the Tables \ref{NGC2527}$-$\ref{NGC2482} for
the number of 
lines used for each star) for which reliable estimates of $gf$-values are available and
the usage of the same lines will remove the systematic uncertainties.

\begin{table*}
 \begin{minipage}{125mm}
 \caption{Comparison of elemental abundance ratios [X/Fe] for open cluster M 67
using data from 
          various sources.}
 \label{compare}
 \begin{tabular}{lccccc}
  \hline
\multicolumn{1}{l}{Species} & \multicolumn{1}{c}{This study}
& \multicolumn{1}{l}{Tautvai{\v s}iene00} &\multicolumn{1}{c}{Yong05} &
\multicolumn{1}{c}{Pancino10}  & \multicolumn{1}{c}{Friel10}  \\
\hline

$[$Na {\scs I}/Fe$]$  &$+0.25\pm0.02$ &$+0.20\pm0.00$ &$+0.30\pm0.05$ &$+0.08\pm0.09$
&$+0.13\pm0.06$  \\
$[$Mg {\scs I}/Fe$]$  &$+0.16\pm0.02$ &$+0.09\pm0.00$ &$+0.16\pm0.05$ &$+0.27\pm0.04$
&$+0.05\pm0.06$  \\
$[$Al {\scs I}/Fe$]$  &$+0.09\pm0.01$ &$+0.13\pm0.02$ &$+0.17\pm0.03$ &$+0.03\pm0.02$
&$+0.11\pm0.04$  \\
$[$Si {\scs I}/Fe$]$  &$+0.20\pm0.02$ &$+0.09\pm0.04$ &$+0.09\pm0.06$ &$+0.10\pm0.02$
&$+0.18\pm0.06$  \\
$[$Ca {\scs I}/Fe$]$  &$+0.04\pm0.02$ &$+0.06\pm0.10$ &$+0.07\pm0.02$ &$-0.16\pm0.03$
&$-0.08\pm0.08$  \\
$[$Ti {\scs I}/Fe$]$  &$-0.01\pm0.02$ &$+0.03\pm0.10$ &$+0.12\pm0.02$ &$-0.04\pm0.06$
&$-0.14\pm0.06$  \\
$[$Cr {\scs I}/Fe$]$  &$+0.05\pm0.02$ &$+0.07\pm0.09$ &$\ldots$       &$+0.01\pm0.03$
&$+0.00\pm0.02$  \\ 
$[$Fe {\scs I}/H$]$   &$-0.08\pm0.02$ &$ 0.00\pm0.06$ &$+0.02\pm0.08$ &$+0.05\pm0.02$
&$+0.03\pm0.08$  \\
$[$Ni {\scs I}/Fe$]$  &$+0.10\pm0.02$ &$+0.06\pm0.09$ &$+0.08\pm0.05$ &$+0.05\pm0.01$
&$-0.02\pm0.05$  \\ 
$[$Y {\scs II}/Fe$]$  &$+0.03\pm0.02$ &$-0.02\pm0.11$ &$\ldots$       &$-0.05\pm0.04$
&$\ldots$      \\
$[$Zr {\scs I}/Fe$]$  &$-0.07\pm0.02$ &$-0.19\pm0.09$ &$-0.28\pm0.02$ &$\ldots$      
&$-0.14\pm0.03$  \\
$[$Ba {\scs II}/Fe$]$ &$\bf-0.16\pm0.06$ &$+0.08\pm0.00$ &$\bf-0.02\pm0.05$ &$+0.25\pm0.02$ 
&$\ldots$      \\
$[$La {\scs II}/Fe$]$ &$ 0.00\pm0.02$ &$+0.12\pm0.05$ &$+0.11\pm0.03$ &$+0.05\pm0.06$
&$\ldots$    \\
$[$Ce {\scs II}/Fe$]$ &$\bf-0.02\pm0.03$ &$+0.08\pm0.08$  &$\ldots$       &$\ldots$
&$\ldots$    \\  
$[$Nd {\scs II}/Fe$]$ &$+0.02\pm0.03$  &$\ldots$      &$\ldots$       &$\ldots$  
&$\ldots$    \\
$[$Sm {\scs II}/Fe$]$ &$\bf-0.03\pm0.03$ &$+0.06\pm0.00$   &$\ldots$   &$\ldots$ 
&$\ldots$    \\
$[$Eu {\scs II}/Fe$]$ &$\bf+0.08\pm0.03$ &$\bf+0.07\pm0.00$ &$\bf+0.06\pm0.02$ &$\ldots$
&$\ldots$    \\

\hline
\end{tabular}
 \flushbottom{{\bf Note}: Abundances calculated by synthesis are presented in bold
numbers and the quoted errors on them are the sensitivity of their abundance to
adopted model atmospheric parameters. An internal uncertainty of 0.00 means that
the abundance analysis is based on a single line only.}
\begin{flushleft}
 Tautvai{\v s}iene00$-$ Tautvai{\v s}iene et al. (2000); Yong05$-$ Yong et al. (2005);
 Pancino10$-$ Pancino et al. (2010); Friel10$-$ Friel et al. (2010);          
\end{flushleft}
\end{minipage}
\end{table*}

A comparison of our observed [Rb/Zr] abundance ratios for these OCs (Figure \ref{rbzr})
 with the recent models
of AGB nucleosynthesis from Figure 14 in Smith et al. (2000) suggests that the 3
M$_{\odot}$ models provide 
complete overlap with observations except for the [Rb/Zr] ratio in the OC NGC 2266.
For these low mass AGB stars, the dominant neutron source is
$^{13}$C($\alpha$,n)$^{16}$O reaction 
which is responsible for the production of s-process elements to observed values. 
We note that NGC 2266 
has a higher [Rb/Zr] ratio than the other OCs studied here, although our measured Rb
abundance is based on 
a single star. Such a high [Rb/Zr] ratio for the OC NGC 2266, complimented with
sub-solar s-process 
elemental abundances and over-solar r-process and $\alpha$-elements, reflects the
addition of 
Tyepe II SNe to the chemical enrichment of its parent gas cloud prior to its
formation. 

Our present sample includes the well-studied OC NGC 2682 (M 67) for which
high-resolution spectroscopic abundance analyses have been reported by 
Tautvai{\v s}iene et al. (2000, hereafter Tautvai{\v s}iene00), 
Yong et al. (2005, hereafter Yong05), 
Pancino et al. (2010, hereafter Pancino10) and Friel et al. (2010, hereafter Friel10).  
Unfortunately, none of the above groups, except Tautvai{\v s}iene00 sample, have 
stars in common with our sample for a direct star-to-star comparison of
results, but on
the assumption that the intrinsic dispersion in abundances within NGC 2682 is
negligible, we may compare results from the separate analyses. For the
Tautvai{\v s}iene00 sample we give in Table \ref{compare} the mean of the average 
abundances of three stars in common with our study and this mean value does not
change if we consider the whole sample of stars studied by Tautvai{\v s}iene00.   
Differences in [X/Fe] between our results and the four published analyses are
$\pm$0.15 dex
or smaller for almost all elements in common: exceptions include Si and Zr for Yong05,
Na, Ca, and Ba for Pancino10, Ba for Tautvai{\v s}iene00, but all elements considered 
by Friel10 fall within the $\pm$0.15 limit. Further, a direct star-to-star 
comparison of [X/Fe] abundance for almost all elements in common between ours and 
Tautvai{\v s}iene00 fall within $\pm$0.15 dex with the exception of [Ba/Fe]. 
The [Fe/H] estimates for the four published analyses
are consistent but our result is consistently slightly sub-solar: [Fe/H] = $-0.08$ versus
$ 0.00$ to $+0.05$. This comparison (Table \ref{compare}) suggests  
a quite satisfactory agreement for all elements given the many different choices which
enter into a LTE model atmosphere abundance analysis. 

It seems very likely that
a concerted effort to establish a truly common method of analysis for red giants in
OCs for use by all the principal players should minimize or even
eliminate systematic abundance
differences. Then, a combined set of results for OCs across the Galaxy should
reveal novel insights into the development of abundance differences in time and
space within the Galactic disc.

\section{Conclusions}

This study is a part of a project devoted to the measurement of metallicity
and chemical abundances of giant stars in OCs covering a broad range of elements,
since such measurements are often lacking in the literature, especially for
neutron capture elements.
We have performed a detailed differential abundance analysis of red giants in seven OCs
located in the Galactic anticentre direction at Galactocentric distances of R$_{gc}$
$\sim$
8.3 to 11.3 kpc. Based on the high resolution spectra and standard LTE analysis, 
we have derived  stellar parameters and abundance ratios for  elements
from Na to Eu. 
 
The main results of this study are as follows: 

\begin{enumerate}

\item We have derived average [Fe/H] values of $-0.10\pm0.04$ for NGC 2527, 
  $-0.08\pm0.04$ for NGC 2682, $-0.07\pm0.04$ for NGC 2482, $-0.06\pm0.04$ for NGC
 2539, $-0.18\pm0.03$ for NGC 2335, $-0.10\pm0.05$ for NGC 2251, and $-0.44\pm$0.04 for
 NGC 2266. 

\item Our findings confirm the result from Paper {\scs I}
 that these OC and local disc field giants with [Fe/H] $\sim$ 0 have
 identical compositions to within the errors of measurements. 
 The observed offset of [Mn/Fe] between the Luck \& Heiter's (2007) 
 sample of field giants and the field dwarfs/OC giants might be arising 
 due to the neglect of hfs treatment in their abundance analysis of field giants.

\end{enumerate}

\vskip1ex 
\noindent
{\bf Acknowledgments:}

We are greatful to the McDonald Observatory's Time Allocation Committee for
granting us observing time to this project. DLL wishes to thank the Robert 
A. Welch Foundation of Houston, Texas for support through grant F-634. We 
also thank the anonymous referee for useful comments which led to improvements of the
paper.

This research has made use of the 
WEBDA database, operated at the Institute for Astronomy of the University of Vienna 
and the NASA ADS, USA. This research has also made use of Aladin. 
This publication makes use of data products from the Two Micron All Sky Survey, 
which is a joint project of the University of Massachusetts and the Infrared 
Processing and Analysis Center/California Institute of Technology, funded by 
the National Aeronautics and Space Administration (NASA) and the 
National Science Foundation (NSF).

\newpage

\begin{table*}
\centering
 \begin{minipage}{90mm}
 \caption{Elemental abundances for stars in the OC NGC 2527.}
 \label{NGC2527}
 \begin{tabular}{llll}
  \hline
\multicolumn{1}{c}{Species}& \multicolumn{1}{c}{star no. 10}& 
\multicolumn{1}{c}{star no. 203} & \multicolumn{1}{c}{Average} \\
\hline

$[$Na I/Fe$]$ &$+0.31\pm0.03$(5) &$+0.33\pm0.03$(4)  & $+0.32\pm0.02$  \\ 
$[$Mg I/Fe$]$ &$+0.07\pm0.01$(4) &$+0.08\pm0.02$(5)  & $+0.07\pm0.01$  \\
$[$Al I/Fe$]$ &$+0.06\pm0.04$(6) &$+0.04\pm0.03$(4)  &$+0.05\pm0.02$   \\
$[$Si I/Fe$]$ &$+0.20\pm0.03$(13)&$+0.21\pm0.03$(10) &$+0.20\pm0.02$   \\
$[$Ca I/Fe$]$ &$+0.09\pm0.05$(14) &$+0.15\pm0.04$(11)  &$+0.12\pm0.03$ \\
$[$Sc I/Fe$]$ &$+0.30\pm0.06$(6)  &$+0.21\pm0.04$(7)   &$+0.25\pm0.04$\\
$[$Sc II/Fe$]$&$+0.15\pm0.05$(5)  &$+0.07\pm0.04$(5)   &$+0.12\pm0.03$  \\
$[$Ti I/Fe$]$ &$+0.12\pm0.02$(13)&$+0.11\pm0.04$(16)   &$+0.11\pm0.02$  \\
$[$Ti II/Fe$]$&$+0.08\pm0.04$(6) &$+0.09\pm0.03$(7)    &$+0.08\pm0.02$ \\
$[$V I/Fe$]$  &$+0.24\pm0.03$(14)&$+0.18\pm0.05$(16)   &$+0.21\pm0.03$ \\
$[$Cr I/Fe$]$ &$+0.11\pm0.04$(9) &$+0.10\pm0.04$(10)  &$+0.10\pm0.03$  \\ 
$[$Cr II/Fe$]$&$+0.09\pm0.04$(9) &$+0.03\pm0.05$(7)   &$+0.06\pm0.03$   \\
$[$Mn I/Fe$]$ & $\bf-0.03 $          &$\bf+0.04$      &$\bf 0.00$    \\
$[$Fe I/H$]$  &$-0.14\pm0.04$(47)&$-0.09\pm0.04$(55)  &$-0.11\pm0.03$  \\ 
$[$Fe II/H$]$ &$-0.11\pm0.04$(13)&$-0.07\pm0.03$(12)  &$-0.09\pm0.02$  \\
$[$Co I/Fe$]$ &$+0.21\pm0.04$(4) &$+0.11\pm0.06$(4)   &$+0.16\pm0.04$  \\
$[$Ni I/Fe$]$ &$+0.05\pm0.04$(18)&$+0.08\pm0.03$(17)  &$+0.06\pm0.02$   \\
$[$Cu I/Fe$]$ &    $\bf-0.11$    &  $\bf-0.18$        &$\bf-0.14$    \\
$[$Zn I/Fe$]$ &    $\bf-0.17$    &  $\bf-0.16$        &$\bf-0.16$     \\
$[$Rb I/Fe$]$ &    $\bf+0.11$    &  $\bf+0.03$        &$\bf+0.07$  \\
$[$Y II/Fe$]$ &$+0.17\pm0.03$(5) &$+0.15\pm0.03$(5)   &$+0.16\pm0.02$  \\  
$[$Zr I/Fe$]$ &$+0.38\pm0.07$(5) &$+0.25\pm0.07$(4)   &$+0.31\pm0.05$  \\
$[$Zr II/Fe$]$&$+0.26\pm0.02$(3) &$+0.10\pm0.04$(2)   &$+0.18\pm0.02$  \\
$[$Ba II/Fe$]$&  $\bf+0.09$      & $\bf+0.07$         &$\bf+0.08$   \\
$[$La II/Fe$]$&$+0.26\pm0.04$(3) &$+0.27\pm0.03$(3)   &$+0.26\pm0.02$ \\ 
$[$Ce II/Fe$]$& $+0.26$          &$\bf+0.22$          &$\bf+0.24$ \\
$[$Nd II/Fe$]$&$+0.20\pm0.01$(4) &$+0.20\pm0.04$(6)   &$+0.20\pm0.02$ \\ 
$[$Sm II/Fe$]$&  $\bf+0.19$      & $\bf+0.18$         &$\bf+0.18 $     \\
$[$Eu II/Fe$]$&  $\bf+0.12$      & $\bf+0.08$         &$\bf+0.10$   \\

\hline
 \end{tabular}
\vskip1ex
\flushbottom{{\bf Note}: The abundances calculated by synthesis are presented in
bold numbers.
The remaining elemental abundances were calculated using line equivalent widths.
Numbers in the parentheses indicate the number of lines used in calculating 
the abundance of that element.
In this analysis we have adopted the hfs data of Prochaska \& McWilliam (2000) for
Mn I, Mucciarelli et al. (2008) for Eu II line, McWilliam (1998) 
for Ba II line, and Allen et al. (2011) for Cu I lines.
}
\end{minipage}
\end{table*}

\begin{table*}
\centering
 \begin{minipage}{90mm}
 \caption{Elemental abundances for stars in the OC NGC 2682.}
 \label{NGC2682}
 \begin{tabular}{lllll}
  \hline
\multicolumn{1}{c}{Species}& \multicolumn{1}{c}{star no. 84}& 
\multicolumn{1}{c}{star no. 151} & \multicolumn{1}{c}{star no. 164}
&\multicolumn{1}{c}{Average} \\
\hline

$[$Na I/Fe$]$ &$+0.23\pm0.05$(6) &$+0.27\pm0.02$(6) &$+0.26\pm0.02$(6)
&$+0.25\pm0.02$  \\
$[$Mg I/Fe$]$ &$+0.12\pm0.04$(7) &$+0.17\pm0.04$(8) &$+0.20\pm0.04$(8)
&$+0.16\pm0.02$   \\
$[$Al I/Fe$]$ &$+0.09\pm0.02$(5) &$+0.05\pm0.03$(4) &$+0.12\pm0.03$(6)
&$+0.09\pm0.01$   \\
$[$Si I/Fe$]$
&$+0.19\pm0.04$(13)&$+0.18\pm0.05$(14)&$+0.23\pm0.03$(15)&$+0.20\pm0.02$   \\
$[$Ca I/Fe$]$ &$+0.06\pm0.03$(14)&$+0.02\pm0.05$(13)&$+0.03\pm0.03$(15)
&$+0.04\pm0.02$   \\
$[$Sc I/Fe$]$ &$+0.11\pm0.04$(4) &$+0.01\pm0.05$(6) &$-0.01\pm0.06$(7) 
&$+0.04\pm0.03$   \\
$[$Sc II/Fe$]$&$+0.20\pm0.01$(4) &$+0.09\pm0.06$(3) &$ 0.00\pm0.04$(2) 
&$+0.10\pm0.02$  \\ 
$[$Ti I/Fe$]$&$+0.06\pm0.04$(18) &$-0.04\pm0.04$(19)&$-0.06\pm0.04$(18) 
&$-0.01\pm0.02$  \\
$[$Ti II/Fe$]$&$+0.08\pm0.02$(9) &$-0.01\pm0.04$(9) &$-0.03\pm0.02$(7)  
&$+0.01\pm0.02$  \\
$[$V I/Fe$]$  &$+0.13\pm0.05$(13)&$+0.08\pm0.04$(12)&$+0.06\pm0.04$(14) 
&$+0.09\pm0.02$  \\
$[$Cr I/Fe$]$ &$+0.08\pm0.03$(14) &$+0.02\pm0.04$(18)&$+0.05\pm0.03$(14)
&$+0.05\pm0.02$   \\ 
$[$Cr II/Fe$]$&$+0.09\pm0.02$(9)  &$+0.06\pm0.05$(10)&$+0.10\pm0.02$(10)
&$+0.08\pm0.02$   \\ 
$[$Mn I/Fe$]$ &$\bf -0.08 $   &$\bf-0.12 $    &$\bf-0.03 $  &$\bf-0.08$  
   \\
$[$Fe I/H$]$  &$-0.08\pm0.03$(66)&$-0.06\pm0.04$(64)&$-0.11\pm0.04$(64)
&$-0.08\pm0.02$  \\
$[$Fe II/H$]$ &$-0.07\pm0.04$(11)&$-0.06\pm0.04$(10)&$-0.11\pm0.03$(11)
&$-0.08\pm0.02$   \\ 
$[$Co I/Fe$]$ &$+0.14\pm0.04$(4) &$+0.06\pm0.05$(4) &$+0.13\pm0.02$(5) 
&$+0.11\pm0.02$  \\ 
$[$Ni I/Fe$]$ &$+0.11\pm0.03$(20)&$+0.08\pm0.03$(21)&$+0.10\pm0.04$(20)
&$+0.10\pm0.02$   \\ 
$[$Cu I/Fe$]$ &   $\bf+0.05$    &  $\bf+0.01$    &$\bf-0.16$  & $\bf-0.03$    
 \\
$[$Zn I/Fe$]$ &   $\bf+0.04$    &  $\bf-0.12$    &$\bf-0.14$  & $\bf-0.07$    
 \\ 
$[$Rb I/Fe$]$ &   $\bf-0.10$   &  $\bf-0.19$     &$\bf-0.11$  &$\bf-0.10$    
  \\
$[$Y II/Fe$]$ &$+0.07\pm0.04$(5) &$-0.01\pm0.02$(6) &$+0.04\pm0.02$(7) 
&$+0.03\pm0.02$   \\
$[$Zr I/Fe$]$ &$ 0.00\pm0.04$(7) &$-0.11\pm0.03$(7) &$-0.11\pm0.05$(7) 
&$-0.07\pm0.02$  \\
$[$Zr II/Fe$]$&$-0.01\pm0.03$(2) &$-0.07\pm0.05$(3) &$-0.13\pm0.00$(1)  
&$-0.07\pm0.03$  \\
$[$Ba II/Fe$]$&  $\bf-0.16 $     &  $\bf-0.16$     &$\bf-0.17$      &$\bf-0.16$     
 \\
$[$La II/Fe$]$&$+0.09\pm0.03$(6) &$-0.03\pm0.04$(6) &$-0.07\pm0.02$(6) &$
0.00\pm0.02$   \\
$[$Ce II/Fe$]$&$\bf-0.09$         &$\bf+0.01$        &$\bf+0.02$    &$\bf-0.02$      \\
$[$Nd II/Fe$]$&$+0.07\pm0.07$(7) &$-0.02\pm0.03$(10) &$
0.00\pm0.03$(10)&$+0.02\pm0.03$   \\
$[$Sm II/Fe$]$&  $\bf+0.02 $        &$\bf -0.07 $    &$\bf-0.04 $   &$\bf-0.03$    \\
$[$Eu II/Fe$]$&  $\bf+0.12 $        & $\bf+0.07$     &$\bf+0.05$    &$\bf+0.08$    \\

\hline
 \end{tabular}
\vskip1ex
\flushbottom{{\bf Note}:Same as in table \ref{NGC2527}.}
\end{minipage}
\end{table*} 

\newpage

\begin{table*}
\centering
 \begin{minipage}{90mm}
 \caption{Elemental abundances for stars in the OC NGC 2539.}
 \label{NGC2539}
 \begin{tabular}{llll}
  \hline
\multicolumn{1}{c}{Species} & \multicolumn{1}{c}{star no. 346} & 
\multicolumn{1}{c}{star no. 463} & \multicolumn{1}{c}{Average} \\
\hline

$[$Na I/Fe$]$ &$+0.27\pm0.01$(4) &$+0.27\pm0.04$(4)  & $+0.27\pm0.02$   \\
$[$Mg I/Fe$]$ &$+0.06\pm0.05$(5) &$+0.08\pm0.02$(4)  & $+0.07\pm0.02$   \\
$[$Al I/Fe$]$ &$-0.05\pm0.02$(5) &$+0.05\pm0.01$(4)  &$ 0.00\pm0.01$   \\
$[$Si I/Fe$]$ &$+0.14\pm0.03$(12)&$+0.21\pm0.03$(12) &$+0.17\pm0.02$   \\
$[$Ca I/Fe$]$ &$+0.02\pm0.04$(12) &$+0.06\pm0.04$(15)  &$+0.04\pm0.03$  \\
$[$Sc I/Fe$]$ &$+0.21\pm0.03$(7)  &$+0.21\pm0.04$(5)   &$+0.21\pm0.02$   \\
$[$Sc II/Fe$]$&$+0.05\pm0.06$(5)  &$+0.03\pm0.05$(5)   &$+0.04\pm0.04$   \\ 
$[$Ti I/Fe$]$ &$+0.10\pm0.04$(15)&$+0.11\pm0.04$(14)   &$+0.10\pm0.03$  \\
$[$Ti II/Fe$]$&$+0.08\pm0.03$(7) &$+0.08\pm0.05$(6)    &$+0.08\pm0.03$   \\
$[$V I/Fe$]$  &$+0.18\pm0.05$(13)&$+0.14\pm0.04$(13)   &$+0.16\pm0.03$   \\
$[$Cr I/Fe$]$ &$+0.10\pm0.03$(10)&$+0.07\pm0.04$(8)   &$+0.08\pm0.02$   \\ 
$[$Cr II/Fe$]$&$+0.09\pm0.02$(6) &$+0.07\pm0.03$(6)   &$+0.08\pm0.02$    \\ 
$[$Mn I/Fe$]$ &$\bf0.00 $        &$\bf+0.02$          &$\bf+0.01$       \\
$[$Fe I/H$]$  &$-0.07\pm0.04$(51)&$-0.05\pm0.04$(66)  &$-0.06\pm0.03$   \\
$[$Fe II/H$]$ &$-0.07\pm0.03$(13)&$-0.07\pm0.04$(14)  &$-0.07\pm0.02$   \\ 
$[$Co I/Fe$]$ &$+0.13\pm0.04$(4) &$+0.01\pm0.04$(6)   &$+0.07\pm0.03$   \\ 
$[$Ni I/Fe$]$ &$ 0.00\pm0.04$(17)&$+0.05\pm0.03$(21)  &$+0.02\pm0.02$   \\ 
$[$Cu I/Fe$]$ &  $\bf-0.18$      &  $\bf-0.15$        &$\bf-0.16$        \\
$[$Zn I/Fe$]$ &  $\bf-0.23$      &  $\bf-0.22$        &$\bf-0.22$        \\ 
$[$Rb I/Fe$]$ &  $\bf+0.07$      &  $\bf+0.01$        &$\bf+0.04$      \\
$[$Y II/Fe$]$ &$+0.15\pm0.05$(6) &$+0.19\pm0.04$(5)   &$+0.17\pm0.03$   \\
$[$Zr I/Fe$]$ &$+0.23\pm0.07$(5) &$+0.19\pm0.03$(5)   &$+0.21\pm0.04$   \\
$[$Zr II/Fe$]$&$+0.28\pm0.04$(2) &$+0.25\pm0.03$(2)   &$+0.26\pm0.02$  \\
$[$Ba II/Fe$]$&  $\bf+0.13$      &  $\bf+0.08$        &$\bf+0.10$       \\
$[$La II/Fe$]$&$+0.18\pm0.02$(4) &$+0.18\pm0.04$(5)   &$+0.18\pm0.02$   \\
$[$Ce II/Fe$]$&$\bf+0.21 $       &$\bf+0.20 $         &$\bf+0.20$       \\
$[$Nd II/Fe$]$&$+0.22\pm0.05$(5) &$+0.24\pm0.03$(5)   &$+0.23\pm0.03$   \\
$[$Sm II/Fe$]$&  $\bf+0.16$      & $\bf+0.20$         &$\bf+0.18 $    \\
$[$Eu II/Fe$]$&  $\bf+0.19$      & $\bf+0.19$         &$\bf+0.19$     \\
         
\hline
 \end{tabular}
 \vskip1ex
\flushbottom{{\bf Note}:Same as in table \ref{NGC2527}.}
\end{minipage}
\end{table*}

\begin{table*}
\centering
 \begin{minipage}{90mm}
 \caption{Elemental abundances for stars in the OC NGC 2251.}
 \label{NGC2251}
 \begin{tabular}{llll}
  \hline
\multicolumn{1}{c}{Species} & \multicolumn{1}{c}{star no. 3} & 
\multicolumn{1}{c}{star no. 33} & \multicolumn{1}{c}{Average} \\
\hline

$[$Na I/Fe$]$ &$+0.36\pm0.04$(4) &$+0.31\pm0.04$(4) &$+0.33\pm0.03$  \\
$[$Mg I/Fe$]$ &$+0.10\pm0.05$(6) &$+0.02\pm0.05$(5) &$+0.06\pm0.03$  \\
$[$Al I/Fe$]$ &$+0.03\pm0.04$(5) &$-0.02\pm0.03$(5) &$ 0.00\pm0.02$  \\
$[$Si I/Fe$]$ &$+0.21\pm0.01$(14)&$+0.25\pm0.05$(14)&$+0.23\pm0.02$  \\
$[$Ca I/Fe$]$ &$+0.10\pm0.04$(16) &$+0.08\pm0.03$(13)&$+0.09\pm0.02$ \\
$[$Sc I/Fe$]$ &$+0.09\pm0.03$(5) &$-0.01\pm0.03$(3)  &$+0.04\pm0.02$ \\
$[$Sc II/Fe$]$&$\bf+0.01$        &$\bf+0.03$         &$\bf+0.02$     \\
$[$Ti I/Fe$]$&$-0.01\pm0.04$(18) &$-0.07\pm0.04$(15) &$-0.04\pm0.03$  \\
$[$Ti II/Fe$]$&$-0.06\pm0.04$(8) &$-0.02\pm0.05$(4)  &$-0.04\pm0.03$  \\
$[$V I/Fe$]$  &$-0.01\pm0.04$(16)&$-0.10\pm0.03$(13) &$-0.05\pm0.02$  \\
$[$Cr I/Fe$]$ &$+0.08\pm0.05$(10)&$+0.02\pm0.05$(8)  &$+0.05\pm0.03$  \\ 
$[$Cr II/Fe$]$&$ 0.00\pm0.05$(8) &$+0.08\pm0.05$(7)  &$+0.04\pm0.03$  \\ 
$[$Mn I/Fe$]$ &$\bf-0.12 $       &$\bf-0.14 $        &$\bf-0.13$     \\
$[$Fe I/H$]$  &$-0.11\pm0.04$(56)&$-0.09\pm0.05$(57) &$-0.10\pm0.03$  \\
$[$Fe II/H$]$ &$-0.11\pm0.04$(11)&$-0.09\pm0.04$(12) &$-0.10\pm0.03$  \\ 
$[$Co I/Fe$]$ &$+0.08\pm0.05$(6) &$-0.01\pm0.04$(5)  &$+0.03\pm0.03$  \\ 
$[$Ni I/Fe$]$ &$+0.05\pm0.04$(22)&$+0.04\pm0.04$(17) &$+0.04\pm0.03$  \\ 
$[$Cu I/Fe$]$ &   $\bf-0.19$     &  $\bf-0.27$       &$\bf-0.22$     \\
$[$Zn I/Fe$]$ &   $\bf-0.11$     &  $\bf-0.16$       &$\bf-0.13$     \\ 
$[$Rb I/Fe$]$ &   $\bf-0.17$     &  $\bf-0.18$       &$\bf-0.17$     \\
$[$Y II/Fe$]$ &$+0.05\pm0.03$(5) &$+0.10\pm0.05$(5)  &$+0.07\pm0.03$  \\
$[$Zr I/Fe$]$ &$+0.03\pm0.05$(5) &$-0.02\pm0.03$(4)  &$ 0.00\pm0.03$  \\
$[$Zr II/Fe$]$&$+0.06\pm0.01$(2) &$+0.06\pm0.02$(2)  &$+0.06\pm0.01$  \\
$[$Ba II/Fe$]$&  $\bf+0.12$      &  $\bf+0.10$       &$\bf+0.11$      \\
$[$La II/Fe$]$&$+0.05\pm0.04$(5) &$ 0.00\pm0.03$(3)  &$+0.02\pm0.02$   \\
$[$Ce II/Fe$]$&$\bf+0.07$        &$\bf+0.08$         &$\bf+0.07$      \\
$[$Nd II/Fe$]$&$+0.08\pm0.03$(9) &$+0.09\pm0.04$(8)  &$+0.08\pm0.02$   \\
$[$Sm II/Fe$]$&  $\bf+0.03$      & $\bf+0.01$        &$\bf+0.02$    \\
$[$Eu II/Fe$]$&  $\bf+0.05$      & $\bf+0.04$        &$\bf+0.04$    \\

\hline
 \end{tabular}
\vskip1ex
\flushbottom{{\bf Note}:Same as in table \ref{NGC2527}.}
\end{minipage}
\end{table*} 
 
\newpage

\begin{table*}
\centering
 \begin{minipage}{90mm}
 \caption{Elemental abundances for stars in the OCs NGC 2482, NGC 2335 and NGC 2266.}
 \label{NGC2482}
 \begin{tabular}{llll}
  \hline
\multicolumn{1}{c}{Species} & \multicolumn{1}{c}{star no. 2482\#9} & 
\multicolumn{1}{c}{star no. 2335\#11} &\multicolumn{1}{c}{star no. 2266\#73} \\
\hline

$[$Na I/Fe$]$  &$+0.30\pm0.03$(6) &$+0.24\pm0.02$(6) &$+0.23\pm0.03$(5) \\
$[$Mg I/Fe$]$  &$+0.13\pm0.02$(4) &$+0.08\pm0.02$(6) &$+0.39\pm0.02$(5) \\
$[$Al I/Fe$]$  &$+0.07\pm0.02$(4) &$-0.02\pm0.02$(6) &$+0.25\pm0.02$(5) \\
$[$Si I/Fe$]$  &$+0.23\pm0.04$(13)  &$+0.10\pm0.02$(15) &$+0.28\pm0.02$(15) \\
$[$Ca I/Fe$]$  &$+0.01\pm0.05$(15)  &$+0.09\pm0.04$(15) &$+0.17\pm0.05$(16) \\
$[$Sc I/Fe$]$  &$+0.12\pm0.05$(6)  &$+0.13\pm0.04$(3) &$+0.31\pm0.05$(3) \\
$[$Sc II/Fe$]$ &$+0.08\pm0.05$(4)  & $\bf+0.16$   &$+0.22\pm0.05$(5) \\ 
$[$Ti I/Fe$]$  &$+0.01\pm0.04$(14) &$+0.17\pm0.04$(14) &$+0.23\pm0.04$(16) \\
$[$Ti II/Fe$]$ &$-0.03\pm0.04$(7)  &$+0.12\pm0.04$(10) &$+0.30\pm0.04$(7) \\
$[$V I/Fe$]$   &$+0.10\pm0.05$(16) &$+0.13\pm0.04$(10) &$+0.20\pm0.06$(17) \\
$[$Cr I/Fe$]$  &$+0.10\pm0.04$(12) &$+0.07\pm0.03$(10) &$+0.09\pm0.03$(12) \\ 
$[$Cr II/Fe$]$ &$+0.09\pm0.05$(7)  &$ 0.00\pm0.04$(8)  &$+0.02\pm0.04$(9) \\ 
$[$Mn I/Fe$]$  &   $\bf-0.11$      & $\bf-0.01$        &$\bf-0.01$     \\
$[$Fe I/H$]$   &$-0.07\pm0.04$(62) &$-0.19\pm$0.04(59) &$-0.45\pm$0.04(88) \\
$[$Fe II/H$]$  &$-0.07\pm0.05$(12) &$-0.17\pm$0.03(12) &$-0.43\pm$0.05(12) \\ 
$[$Co I/Fe$]$  &$+0.11\pm0.03$(7)  &$+0.16\pm$0.03(3)  &$+0.27\pm0.03$(5) \\ 
$[$Ni I/Fe$]$  &$+0.03\pm0.04$(17) &$+0.09\pm$0.03(21) &$+0.09\pm0.03$(20) \\ 
$[$Cu I/Fe$]$  &  $\bf-0.21$    &$\bf-0.15$     &$\bf+0.02$     \\
$[$Zn I/Fe$]$  &  $\bf-0.22$    &$\bf-0.06$     &$\bf 0.00$     \\  
$[$Rb I/Fe$]$  &  $\bf-0.13$    &$\bf+0.08$     &$\bf+0.14$     \\ 
$[$Y II/Fe$]$  &$+0.15\pm0.04$(5) &$+0.12\pm$0.05(5) &$+0.05\pm0.04$(4) \\
$[$Zr I/Fe$]$  &$+0.11\pm0.05$(5) &$+0.06\pm$0.04(2) &$-0.08\pm$0.05(5) \\
$[$Zr II/Fe$]$ &$+0.10\pm0.04$(3)&$+0.01\pm0.02$(2)& $\ldots$   \\
$[$Ba II/Fe$]$ &   $\bf+0.09$   &$\bf+0.25$     &$\bf-0.13$     \\
$[$La II/Fe$]$ &$+0.18\pm0.03$(5) &$+0.29\pm$0.03(4) &$-0.02\pm0.04$(6) \\
$[$Ce II/Fe$]$ &$\bf+0.11 $     &$\bf+0.29$     &$\bf-0.07$     \\  
$[$Nd II/Fe$]$ &$+0.13\pm0.04$(7) &$+0.32\pm$0.03(8) &$+0.13\pm0.04$(4) \\
$[$Sm II/Fe$]$ &$+0.13\pm0.04$(4) &$+0.28\pm$0.03(3) &$\bf+0.11$     \\
$[$Eu~II/Fe$]$ &$\bf+0.07$      &$\bf+0.07$     &$\bf+0.40$     \\

\hline
 \end{tabular}
\vskip1ex
\flushbottom{{\bf Note}:Same as in table \ref{NGC2527}.}
\end{minipage}
\end{table*}

\end{document}